# Aggregation of a macromolecule in a nano cube

Pramod Kumar Mishra
Email: pkmishrabhu@gmail.com
*Department of Physics, DSB Campus, Kumaun University, Nainital (Uttarakhand) India-263002*

**Abstract:** We model a macromolecule as an infinitely long Gaussian semi-flexible polymer chain and the conformations of the chain were realized in the nano cube using a cubic lattice. A modified version of the recursion relations is used to calculate the grand canonical partition function of the chain to investigate distinct thermo-dynamical properties of the nano polymer aggregate than corresponding bulk behaviour of the macromolecule. Our analytical estimates on the thermo-dynamical properties of the macromolecule clearly show that the nano aggregate of the polymer chain has interesting and distinct conformational statistics than its corresponding bulk state, and the method described in the present report can be easily extend to investigate the thermodynamics of the self-avoiding polymer chain in the nano dimensions.

**Key words**: Exact results, Nano cube, Gaussian chain, thermodynamics, recursion relations

## 1. Introduction

There is ample interest among researchers to investigate properties of the materials at the nano length scales, and it is because of the fact that at the nano length scales we may appreciate behaviour of a fairly small number of molecules/atoms; and in addition to it, at the nano length scales the materials have large surface effects than its corresponding bulk state [1-2]. Therefore, it is expected that a many body system may possess distinct properties at the nano length scales than its corresponding bulk structure. There are interesting discussion regarding distinct properties of the materials at the nano length scale [1-3]; and where at least along one direction the sample of the material has size of the order few thousand nano meter. The nano materials are extensively studied to improve properties of the materials for its desired application, and the overall size of the material at the nano length scale is much compact, and due to its compact size there are several materials which show useful properties which one cannot expect in its bulk state [4-5].

There are very few reports on the properties of the polymer materials at the nano length scales [6-8]; and thermo-dynamical aspects of the semi-flexible polymer chain at the nano length scale are not understood well [6-10]. Therefore, we have chosen an infinitely long linear semi-flexible Gaussian polymer chain to investigate thermodynamics of the polymer chain in the nano-cube. We have used lattice model for the Gaussian chain to mimic conformations of the polymer chain in the nano cube using a cubic lattice model [11-13]. A modified version of the recursion relations is used to solve the chosen model of the Gaussian semi-flexible polymer chain for the case when a semi-flexible Gaussian polymer chain is polymerized in the nano dimensions.

The Gaussian polymer models are simple to solve analytically, and proposed study on the thermodynamics of the Gaussian chain may be an initial step to further investigate properties of a self-avoiding polymer chain at the nano dimensions. Thus, further improvement in our understanding about polymeric material may be achieved through such study on the thermodynamics of the Gaussian polymer chain in the possible other nano geometries. Therefore, we have chosen an infinitely long Gaussian polymer chain to investigate fundamental aspects of the nano polymer aggregate using method of recursion relations [11]. The method described in the present report may be easily extend to understand thermodynamics of nano polymer aggregate of a self-avoiding polymer chain for its short and an infinite length.

The manuscript is organized as follows: We describe a lattice model for the Gaussian semi-flexible polymer chain in the nano cube in the section 2, and method of calculations is summarize in the section 3. The results were outlined in the section four and we conclude our discussion in the section 5.

## 2. The Model

A lattice model of the Gaussian polymer chain is extensively used to discuss properties of single polymer chain [11-13], and there are useful reports which are based on the analytical calculations[11-15]. These reports gives us understanding of a macromolecule in the various geometries [11-18]. Therefore, we have chosen a cubic lattice to mimic conformations of an infinitely long Gaussian chain in a nano cube. The nano cube is impenetrable. A walker is allowed to take steps along all possible directions (*i. e.* $\pm x$, $\pm y$ & $\pm z$ directions) in the chosen cube and the size of the cube varies, say, from one monomer to four monomers. A monomer may be of 5Å in length such that we have a cube of the least volume which is equal to $1.25X10^{-28}$ $m^3$. One end of the chain is grafted at the origin of the nano cube and each bend in the chain cost an energy penalty $E_b$ [$=E_0(1-\cos\theta)$, and the value of $\theta$ is equal to 0°, 90° and 180°] and this penalty has been incorporated in the model as the stiffness weight of the semi-flexible chain, and here stiffness weight is shown using symbol $k[=Exp(-\beta E_b)]$, [14-18]. Therefore, we have weight factor 1, $k$ and $k^2$ for the bending angles 0°, 90° and 180°, respectively.

There are only four possible options for the walker to take steps when he/she is moving along the edges and on the corner this option reduces to three; the walker can take steps along any of the five directions when he/she is moving on the surface of the nano cube and the walker can take steps along any of the all the possible six directions when he/she is in the bulk. Since, the walker is restricted to take very limited number of steps in the bulk inside the nano cube than corresponding bulk where each direction has infinite extension. Therefore, it is expected that the Gaussian chain has interesting and distinct properties from its corresponding bulk state. In order to appreciate above said facts, a walk of ten monomers long Gaussian chain is shown schematically in the figure no. 1 and there are six bends, which are also shown for a Gaussian semi-flexible chain's conformation in the figure.

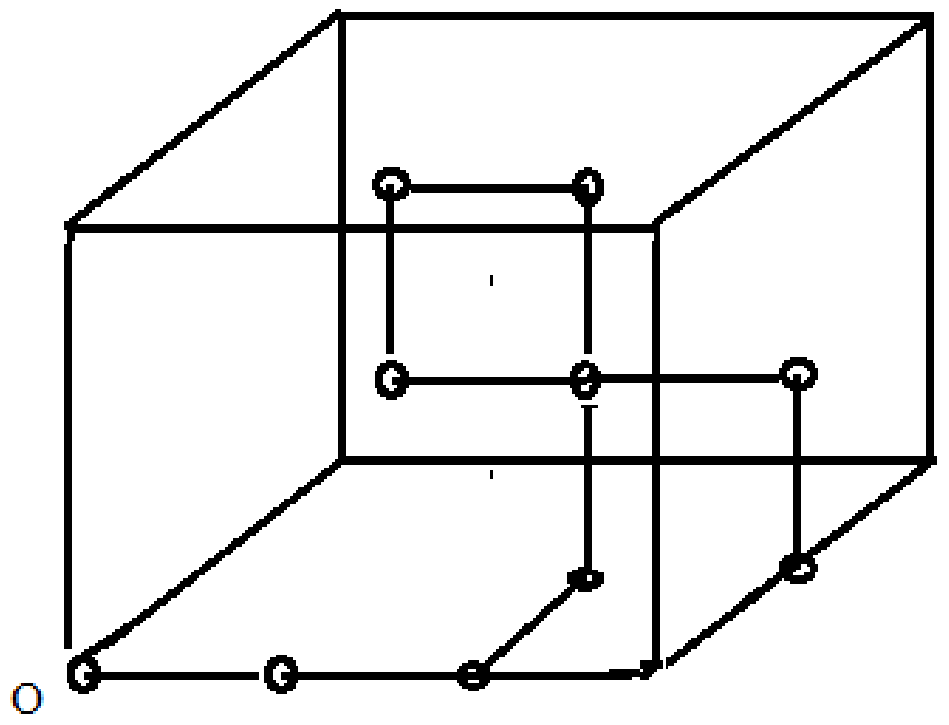

**Figure No. 1:** An impenetrable nano cube is shown schematically and all walks of the chain starts from a point O, and the walker can take steps along all the possible directions in a chosen cubic lattice. We have shown a ten monomers long conformation of the chain and this conformation has six bends.

## 3. Method of Calculations

A general expression for the grand canonical partition function of the Gaussian chain for it's nano cube geometry may be written in the following manner [11,14-18],

$$F(g,k) = \sum_{N=1}^{N \to \infty} \sum_{All-walks-of-N-steps} g^{N} k^{N_b} \tag{1}$$

Few components of the partition function are shown graphically in the figure no. 2 (I) & (II) for the nano cube 3X3X3 units.

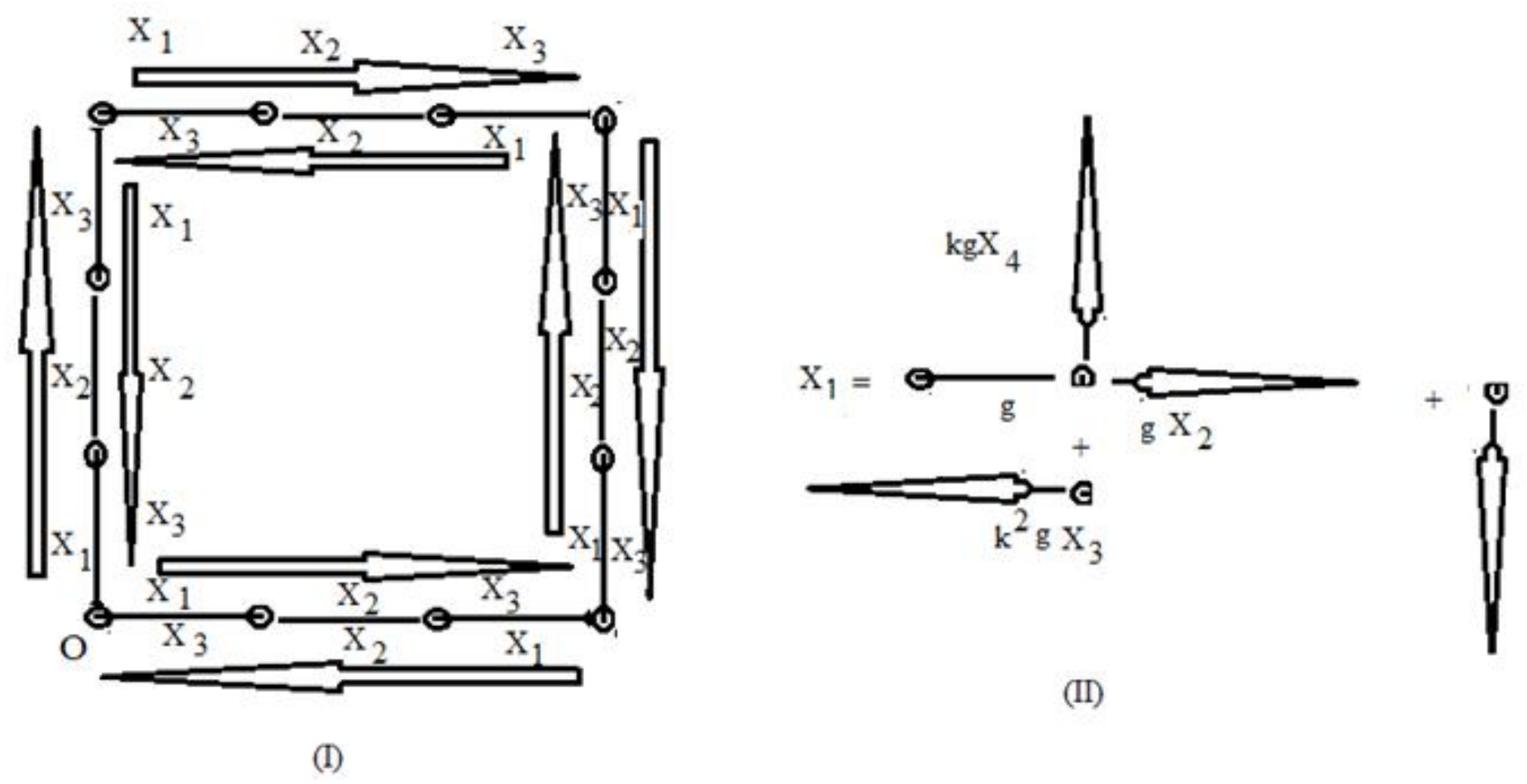

**Figure No. 2:** The components of the partition function along edges of the nano cube are shown graphically in the figure (I) and the impenetrable cube is 3X3X3 units in the volume whose transverse section is shown for the sake of discussion. A component of the grand canonical partition function is shown in the figure (II) for the chosen cube; and here $X_4$ is the component of the partition function on the surface of the cube when the edge of the nano cube is 3 monomers long. Thus, figure no. (II) represents graphically an equation which is written as, $X_1=g+gX_2+k^2gX_3+2kgX_4$ for the nano cube of the volume 3X3X3 unit.

We write components of the recursion relations [11] for a case when the size of the nano cube is 2X2X2, and components are written as follows for the sake of completeness to describe method used in the present manuscript,

$$X_1(g,k) = g + gX_2(g,k) + k^2 gX_2(g,k) + 2kgX_3(g,k) \tag{2}$$

$$X_2(g,k) = g + k^2 gX_1(g,k) + 2kgX_1(g,k) \tag{3}$$

$$X_3(g,k) = g + gX_4(g,k) + k^2 gX_4(g,k) + 2kgX_4(g,k) + kgX_5(g,k) \tag{4}$$

$$X_4(g,k) = g + 2kgX_2(g,k) + k^2 gX_3(g,k) + kgX_3(g,k) \tag{5}$$

$$X_5(g,k) = g + gX_6(g,k) + k^2 gX_6(g,k) + 4kgX_6(g,k) \tag{6}$$

$$X_6(g,k) = g + k^2 gX_5(g,k) + 4kgX_4(g,k) \tag{7}$$

We solved eqn. nos. (2-7) to get the components of the partition function and also the expression for the grand canonical partition function of an infinitely long chain in an impenetrable nano cube of 2X2X2 unit and the partition function is written in the following manner,

$$F_{m=2}(g,k) = 3X_1(g,k) = -\frac{3g(g^5(k-1)^2k^3(k^5+9k^4+26k^3+22k^2-7k-3)+g^4k^3(k^5+5k^4-16k^2+7k+3)-g^3k(2k+1)(k^2+2k-1)^2+g^2(-2k^4-5k^3+2k^2+k)+g(k+1)^2+1)}{g^6(k-1)^2k^4(k^6+11k^5+44k^4+74k^3+37k^2-17k-6)-g^4k^2(3k^6+18k^5+28k^4-13k^2-2k+2)+g^2k(3k^3+9k^2+5k+3)-1} \tag{8}$$

The singularity of the partition function is used to determine the critical value of the monomer fugacity [$g_c(k)$] for the polymerization of an infinitely long semi-flexible chain in the nano cube. The critical value of the monomer fugacity is shown in the figure no. 3 for the nano cube as well as for the chain in the bulk. The partition function of the nano polymer aggregate is used to determine other relevant thermo-dynamical parameters of the chain.

## 4. The results

The bulk corresponds to a situation where extension of the cube is infinite along all six directions (i. e. along ±*x*, ±*y* & ±*z* directions) [11-15]. The nano polymer aggregate has distinct properties than the properties of the chain in the bulk. It is clearly seen from figure no. 3 that the monomer fugacity has distinction from corresponding bulk for the case of a flexible and semi-flexible polymer chain.

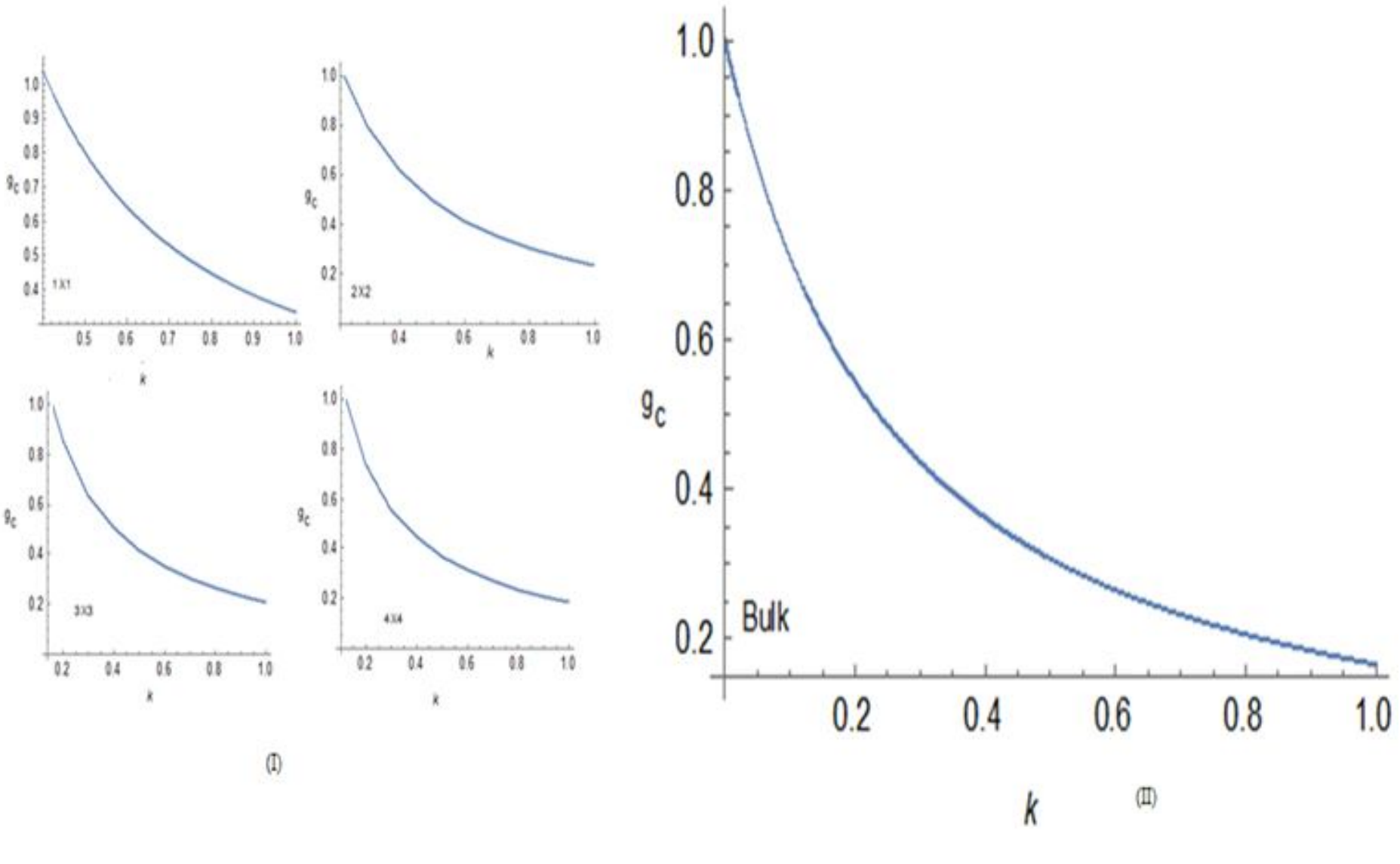

**Figure No. 3:** This figure shows probability of polymerization of an infinitely long semi-flexible Gaussian chain in an impenetrable nano cube (I), and the probability for the polymerization of an infinitely long semi-flexible chain is also shown in the figure no.(II) for the bulk case. This figure shows distinction in the probability of polymerization of an infinitely long chain in the nano cube from the bulk.

The minimum value of the stiffness weight of an infinitely long semi-flexible Gaussian polymer chain for the nano cube is shown in the figure no. 4. It is seen from our analytical calculations that the minimum value of the stiffness weight has value 0.414 for a cube of unit dimensions and stiffness of the chain varies to 0.12 for the nano cube of volume 4X4X4 unit. Thus, a small size of the cube restricts to a possible smallest value of the stiffness weight or the largest finite value of the bending energy of the semi-flexible Gaussian chain which is polymerized in the form of the nano polymer aggregate.

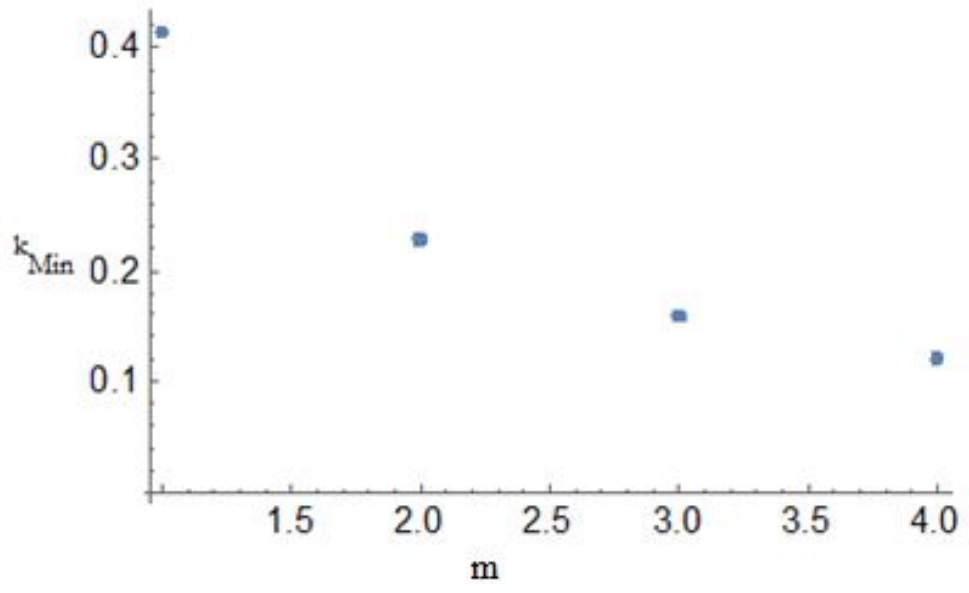

**Figure No.4:** This figure shows possible minimum value of the stiffness of an infinitely long chain in the nano cube. The possible minimum length ($m$) of the nano cube varies from one monomer size to four length and correspondingly the minimum value of the stiffness weight varies from 0.414 to 0.12, respectively.

We calculated average length of the semi-flexible and the flexible polymer chains for its unit bend [13-18], and this parameter is nothing but the persistent length of the Gaussian semi-flexible polymer chain. The persistent length ($l_P$) of the chain diverges for the stiff chain in the bulk, and while in the case of the nano polymer aggregates the persistent length of the semi-flexible polymer chain is less than size ($m$) of the nano cube. The nature of variation of the persistent length of the semi-flexible chain with the stiffness of the chain has different behaviour for the nano polymer aggregate than the polymer chain in the bulk. The details on the persistent length is shown in the figure no. 5.

## 1. The conclusions

We solve lattice model for an infinitely long Gaussian semi-flexible polymer chain in three dimensions using method of recursion relation [11], and our analytical estimates on the thermodynamics of the Gaussian semi-flexible chain in the bulk has different thermodynamics than its corresponding nano polymer aggregate. Such distinction in the Gaussian chain statistics is expected as the monomer fraction on the surface for the nano cube is non zero than corresponding bulk conformations of the chain. The monomer fraction on surface of the polymer chain is zero for an infinitely long chain in the bulk, and therefore, the nano polymer aggregate and the polymer bulk structure are different in respect to thermo-dynamical properties of the polymer chain. It is clearly seen from figure nos. (3), (4) and (5).

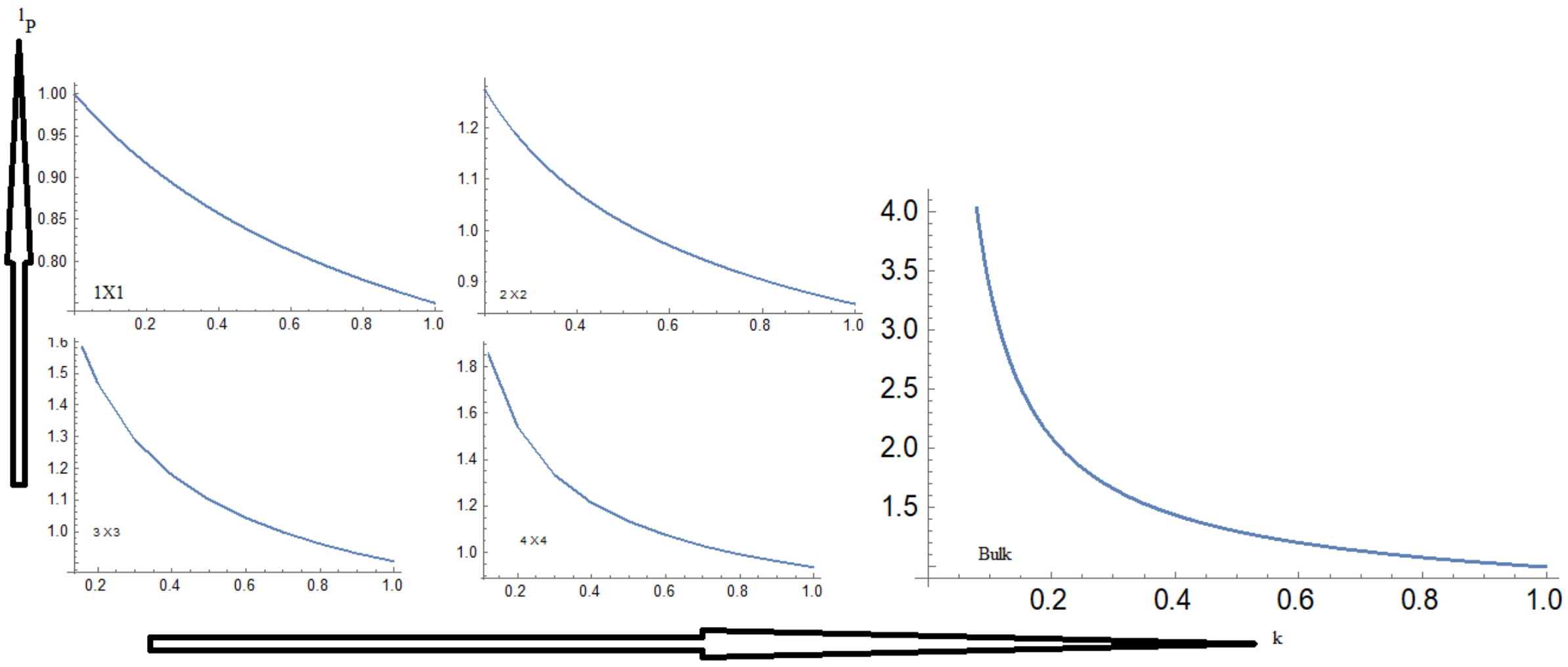

Figure No. 5: We have shown value of the persistent length ($l_P$) of the semi-flexible chain in the nano cube and also in the bulk. The persistent of the stiff chain is finite for the nano polymer aggregate while it is an infinite for the stiff chain in the bulk.

The nano polymer aggregate has non zero value of the stiffness weight while the bending energy of stiff chain may be infinite for the stiff chain in the bulk. Therefore, the bending energy for a possible stiff chain is finite for the nano polymer aggregate while the bending energy is an infinite for the possible stiff chain in the bulk. The nature of variation of the monomer fugacity versus stiffness weight for the nano polymer aggregate is different from than the Gaussian semi-flexible polymer chain in the bulk. The bending energy for the nano polymer aggregate is finite unlike the case of stiff Gaussian polymer chain in the bulk. The minimum value of the stiffness weight for the nano polymer aggregate is shown for chosen length of cube. The minimum value of the stiffness weight asymptotically goes to zero. This nature of variation is shown in the figure no. 4.

The persistent length ($l_P$) of a stiff Guassian chain diverges in the limit $k \to 0$, while the persistent length of nano polymer aggregate is finite and its value is less than length of the nano cube ($m$). The value of the persistent length is shown for chosen stiffness of the semi-flexible Gaussian polymer chain, and it is shown in the figure no. 5. It is seen from figure (5) also that the nano polymer aggregate has different nature of the variation of its persistent length with the stiffness weight of the chain than the persistent length of corresponding bulk structure of the polymer chain.

It is to be noted that method of calculations discussed in the present manuscript may be easily extend to calculate thermodynamics of finite as well an infinite semi-flexible Gaussian chain and semi-flexible self-avoiding polymer chain. Author reported thermodynamics of the semi-flexible polymer aggregate on a two dimensional substrate and the manuscript is under publication [19].